\documentclass[aps,prl,twocolumn,superscriptaddress,longbibliography]{revtex4-2}
\usepackage{graphicx}
\usepackage{amsmath,amssymb}
\usepackage{bm}
\usepackage{dcolumn}
\usepackage{float}
\usepackage[OT1]{fontenc} 
\usepackage{url}
\usepackage{mathrsfs}
\usepackage{slashed,comment}
\usepackage{color}
\usepackage{verbatim}
\usepackage{enumitem}
\usepackage{soul,physics}
\usepackage[driverfallback=dvipdfm]{hyperref}
\hypersetup{pdfpagemode=FullScreen,colorlinks=true,breaklinks,urlcolor=blue,linkcolor=blue,citecolor=blue}

\usepackage{subfigure}
\usepackage{bm}

\usepackage{xcolor}

\usepackage{tikz}
\usepackage{tikz-cd}
\usetikzlibrary{arrows}
\usetikzlibrary{intersections}
\usetikzlibrary{shapes.geometric}
\usetikzlibrary{decorations.pathmorphing, patterns,shapes}
\usetikzlibrary{decorations.markings}

\tikzset{
  mid arrow/.style={postaction={decorate,decoration={
        markings,
        mark=at position .575 with {\arrow[#1]{stealth}}
      }}},
  near arrow/.style={postaction={decorate,decoration={
        markings,
        mark=at position .275 with {\arrow[#1]{stealth}}
      }}},
   far arrow/.style={postaction={decorate,decoration={
        markings,
        mark=at position .800 with {\arrow[#1]{stealth}}
      }}},
}

\begin{document}
  
  \title{Scheme to Detect the Strong-to-weak Symmetry Breaking \\ via Randomized Measurements }

  \author{Ning Sun}
  \affiliation{Department of Physics, Fudan University, Shanghai, 200438, China}

  \author{Pengfei Zhang}
  \thanks{PengfeiZhang.physics@gmail.com}
  \affiliation{Department of Physics, Fudan University, Shanghai, 200438, China}
  \affiliation{State Key Laboratory of Surface Physics, Fudan University, Shanghai, 200438, China}
  \affiliation{Shanghai Qi Zhi Institute, AI Tower, Xuhui District, Shanghai 200232, China}
  \affiliation{Hefei National Laboratory, Hefei 230088, China}

  \author{Lei Feng}
  \thanks{leifeng@fudan.edu.cn}
  \affiliation{Department of Physics, Fudan University, Shanghai, 200438, China}
  \affiliation{State Key Laboratory of Surface Physics, Fudan University, Shanghai, 200438, China}
  \affiliation{Institute for Nanoelectronic devices and Quantum computing, Fudan University, Shanghai, 200438, China}
  \affiliation{Shanghai Key Laboratory of Metasurfaces for Light Manipulation, Shanghai, 200433, China}
  \affiliation{Hefei National Laboratory, Hefei 230088, China}

  \date{\today}

  \begin{abstract}
  Symmetry breaking plays a central role in classifying the phases of quantum many-body systems. Recent developments have highlighted a novel symmetry-breaking pattern, in which the strong symmetry of a density matrix spontaneously breaks to the week symmetry.
  This strong-to-weak symmetry breaking is typically detected using multi-replica correlation functions, such as the R\'enyi-2 correlator. In this letter, we propose a practical protocol for detecting strong-to-weak symmetry breaking in experiments using the randomized measurement toolbox. 
  Our scheme involves collecting the results of random Pauli measurements for (i) the original quantum state and (ii) the quantum state after evolution with the charged operators.
  Based on the measurement results, with a large number of samples, we can obtain an unbiased estimate of the R\'enyi-2 correlator. With a small sample size, we can still provide an alternative approach to estimate the phase boundary to a decent accuracy.
  We perform numerical simulations of Ising chains with all-to-all decoherence as an exemplary demonstration.
  Our result opens the opportunity for the experimental studies of the novel quantum phases in mixed quantum states.
  \end{abstract}
  
  \maketitle

  \emph{ \color{blue}Introduction.--} Recent years have witnessed significant breakthroughs in understanding quantum correlations and entanglement in many-body systems. 
  A central focus of these advances is the multi-replica observables, which offer valuable insights that single-replica calculations cannot capture. 
  Particularly, the investigation of the second R\'enyi entropy in both experiment and theory \cite{Islam:2015aa,Kaufman:2016aa,Brydges:2019aa,Elben:2020aa,Enk:2012aa,Elben:2018aa,Elben:2019aa} has provided crucial insights into the thermalization of isolated quantum systems under unitary evolution \cite{Srednicki:1994aa,Deutsch:1991aa}.
  It shows a clear contrast to systems that exhibit many-body localization \cite{Abanin:2019aa,Alet:2018aa,Altman:2015aa,Nandkishore:2015aa,Schreiber:2015aa,Smith:2016aa,Choi:2016aa}. In parallel, the out-of-time-ordered correlator has emerged as a powerful tool for quantifying the spread of quantum information \cite{Larkin:1969aa,Kitaev:2014aa}
  , and revealed the quantum many-body chaos in generic interacting systems \cite{Sekino:2008aa,Hayden:2007aa,Shenker:2014aa,Stanford:2014aa,Roberts:2015aa,Shenker:2015aa,Maldacena:2016aa}.
  Furthermore, a more refined description is provided by the distribution of operator sizes \cite{Roberts:2015aa,Nahum:2018aa,Hunter-Jones:2018aa,Keyserlingk:2018aa,Khemani:2018aa,Qi:2019aa,Lucas:2019aa,Lucas:2020aa,Chen:2020aa,Roberts:2018aa,Qi:2019ab}, which satisfies a non-trivial self-consistency relation for thermodynamical systems with all-to-all interactions \cite{Zhang:2023aa,Liu:2024aa}.

  
  \begin{figure}[t]
    \centering
    \includegraphics[width=0.85\linewidth]{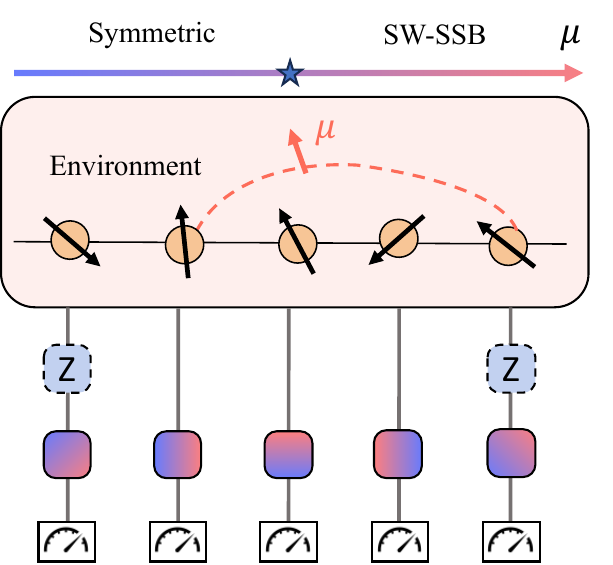}
    \caption{We present a schematic of our protocol for detecting the SW-SSB using randomized measurements. Here, we denote the decoherence strength as $\mu$, which drives a transition between the symmetric phase and the SW-SSB phase. The protocol includes applying $Z$ gates to a pair of distant qubits, before performing random Pauli measurements. A more detailed description of our protocol is provided in the main text.}
    \label{fig:schemticas}
  \end{figure}

  The latest developments further highlight the importance of multi-replica correlation functions in defining phases of mixed states: For a symmetry operation $U$, there are two distinct definitions of symmetric density matrices \cite{Buca:2012aa,Albert:2014aa,Albert:2018aa,Lieu:2020aa}. 
  If the charge in the system and in the bath are conserved separately, the density matrix $\rho$ exhibits strong symmetry $U\rho=e^{i\theta}\rho$. In contrast, for systems that exchange charges with the bath, the density matrix only shows weak symmetry $U\rho~U^\dagger =\rho$. Special attention has been paid to the scenario where the strong symmetry of a density matrix is spontaneously broken to a weak symmetry, called strong-to-weak spontaneous symmetry breaking (SW-SSB) \cite{Lee:2023aa,Ogunnaike:2023aa,Moudgalya:2024aa,Lessa:2024aa,Sala:2024aa,Huang:2024aa,Gu:2024aa,Kuno:2024aa,Zhang:2024aa,Zhang:2024ab,Liu:2024ab,Weinstein:2024aa,Guo:2024aa,Kim:2024aa,Chen:2024aa,Lu:2024aa,Orito:2024aa}. Many studies of the SW-SSB are based on the R\'enyi-2 correlator, defined with two replicas of the density matrix,
  \begin{equation}\label{def:Renyi-2}
  C^{(2)}(j,k)= \frac{\text{tr}[O_jO_k^\dag\rho O_kO_j^\dag\rho]}{\text{tr}[\rho^2]}.
  \end{equation}
  Here, $j,k$ label different qubits, and $O$ is an operator that is charged under the symmetry operation $U$. By definition, the R\'enyi-2 correlator probes the overlap between $\rho$ and $\tilde{\rho}=O_jO_k^\dag\rho O_kO_j^\dag$. The system manifests SW-SSB 
  in the R\'enyi-2 sense
  when $C^{(2)}(j,k)$ shows long-range order while regular correlation functions remain unordered. Despite the importance of the R'enyi-2 correlator, the study of its experimental measurement for the generic density matrix $\rho$ remains lacking.

  In this letter, we propose a concrete protocol for detecting SW-SSB by generalizing the method of the second R\'enyi entropy measurements \cite{Albert:2014aa,Albert:2018aa} using randomized measurement toolbox \cite{Elben:2023aa}. In particular, the numerator of \eqref{def:Renyi-2} can be predicted using the statistical correlation between the results of random Pauli measurements on $\rho$ and $O_jO_k^\dag\rho O_kO_j^\dag$. The validity of our protocol is demonstrated using a quantum Ising model with all-to-all decoherence, introduced as a large-$N$ solvable model for SW-SSB. This model is highly relevant to many state-of-the-art quantum devices, including Rydberg atom arrays \cite{Evered:2023aa,Ma:2023aa,Bluvstein:2024aa,Bekenstein:2020aa,Bluvstein:2021aa,Ebadi:2022aa,Bluvstein:2022aa,Lis:2023aa,Manetsch:2024aa,Tao:2024aa,Cao:2024aa} and trapped ion systems \cite{Monroe:2021aa,Blatt:2012aa,Zhang:2017aa,Joshi:2022aa,Morong:2021aa}. We further discuss the estimation of the phase diagram with a small number of samples, which matches the exact phase diagram with good accuracy. 

  \emph{ \color{blue}The Protocol.--} We consider a quantum many-body system consisting of $N$ qubits with Pauli operators $\{X_i,Y_i,Z_i\}$ ($i=1,2,...,N$). The system is prepared in a
  state with a strongly symmetric density matrix $\rho$, such that $ U\rho=e^{i\theta}\rho$. 
  We aim to propose an experimental protocol to measure $C^{(2)}(j,k)$. In the following discussions, we focus on the Z$_2$ symmetry that $U=\prod_i X_i$ with the charged operators $O_i=Z_i$. Extension to arbitrary unitary operators $O_i$ is straightforward. 
  The denominator of the R\'enyi-2 correlator is the purity $P_I=\text{tr}[\rho^2]$ and is measured using randomized measurements\cite{Brydges:2019aa,Elben:2020aa}. 
  The challenging task is the measurement of the numerator $P_{ZZ}=\text{tr}[Z_jZ_k\rho Z_j Z_k\rho]=\text{tr}[\tilde{\rho}\rho]$. 
  We address this challenge by generalizing the discussions in \cite{Brydges:2019aa,Elben:2020aa,Enk:2012aa,Elben:2018aa,Elben:2019aa}, 
  and propose the measurement scheme with the following steps:
  \begin{enumerate}
  \item Randomly choose a direction $\hat{n}_i\in \{\hat{x},\hat{y},\hat{z}\}$ with equal probability for each qubit, independently;

  \item Prepare the system in the state $\rho$ and measure each qubit $i$ along the direction $\hat{n}_i$. Record the measurement outcome $\bm{s}=s_1s_2...s_N$ with $s_i=\pm 1$; 

  \item Prepare the system in the state $\tilde{\rho}$ by applying additional $Z$ gates to the qubit $j$ and $k$. Then, perform the measure each qubit $i$ along the same direction $\hat{n}_i$ and record the measurement outcome $\bm{s}'=s_1's_2'...s_N'$;

  \item Repeat 1-3 multiple times to collect a sufficiently large dataset $\{\bm{s}_a,\bm{s}'_a\}$, where $a=1,2,...,N_a$ labels the iterations of the experiment. 
  
  \end{enumerate}
  Based on the results of measurements, the numerator can be obtained by computing
  \begin{equation}\label{eqn:relation}
  P_{ZZ}= \lim_{N_a \rightarrow \infty}\sum_{a} 2^{N} (-2)^{-D(\bm{s}_a,\bm{s}'_a)} /N_a,
  \end{equation}
  with $D(\bm{x},\bm{y})$ being the Hamming distance that counts the number of distinct elements between two bit strings.

  We proceed to prove Eq.\eqref{eqn:relation} by replacing the sample average with the expectation value. Thus, proving Eq. \eqref{eqn:relation} becomes equivalent to showing that
  \begin{equation}\label{eqn:relation2}
  \text{tr}[\tilde{\rho} \rho]= 2^{N}\overline{ (-2)^{-D(\bm{s},\bm{s}')}P_{\rho}(\bm{s},\bm{n})P_{\tilde{\rho}}(\bm{s}',\bm{n})}.
  \end{equation}
  Here, we have introduced the probability $P_{\rho}(\bm{s}\bm{n})=(\otimes_i\langle s_i n_i|)\rho(\otimes_i|s_i n_i\rangle)$ under the Born rule. The overline denotes both the average over random directions $\hat{n}_i$ and the experimental results $\bm{s}$ and $\bm{s}'$. 
  
  A key observation is that the above protocol can reduce to the measurement of the purity $P_I=\text{tr}[\rho^2]$ if the $Z$-gate in step 3 is skipped. It also has been established in Ref.~\cite{Brydges:2019aa,Elben:2020aa} and 
  \begin{equation}\label{eqn:Purity}
  \text{tr}[ \hat{\rho}^2]= 2^{N}\overline{ (-2)^{-D(\bm{s},\bm{s}')}P_{\hat\rho}(\bm{s},\bm{n})P_{\hat \rho}(\bm{s}',\bm{n})},
  \end{equation}
  for arbitrary density matrix $\hat{\rho}$. When we take $\hat{\rho}=(\rho+\tilde{\rho})/2$, the L.H.S. of Eq.~\eqref{eqn:Purity} becomes $\text{tr}[\rho^2]/4+\text{tr}[\tilde{\rho}^2]/4+\text{tr}[\tilde{\rho} \rho]/2$. On the R.H.S., we apply the linearity of the probability 
  \begin{equation}
  P_{\hat{\rho}}(\bm{s},\bm{n})=\frac{P_{\rho}(\bm{s},\bm{n})+P_{\tilde\rho}(\bm{s},\bm{n})}{2}.
  \end{equation}
  This leads to three different terms. The two terms involving $P_{\rho}(\bm{s},\bm{n})P_{\rho}(\bm{s}',\bm{n})$ and $P_{\tilde\rho}(\bm{s},\bm{n})P_{\tilde\rho}(\bm{s}',\bm{n})$ again takes the the form of Eq. \eqref{eqn:Purity}, with $\hat{\rho}=\rho$ or $\hat{\rho}=\tilde{\rho}$, respectively. 
  Their contribution only leads to $\text{tr}[\rho^2]/4+\text{tr}[\tilde{\rho}^2]/4$ that cancel the terms on the L.H.S.. The remaining cross term $P_{\rho}(\bm{s},\bm{n})P_{\tilde{\rho}}(\bm{s}',\bm{n})$ meets $\text{tr}[\tilde{\rho} \rho]/2$ and gives Eq.~\eqref{eqn:relation2}, and thus validates our measurement protocol.

  \emph{ \color{blue}The Model.--} 
  To implement our protocol detecting SW-SSB transition, we first consider the solvable model for quantum spin chains under all-to-all decoherence. We take the transverse field Ising model as an example
  \begin{equation}\label{eqn:Ising}
  H=-\sum_{i} Z_iZ_{i+1}-g X_i.
  \end{equation}
  The system is initialized in the ground state $|\psi_0\rangle$. To avoid conventional symmetry breaking, we restrict the discussion to $g>1$, where the $|\psi_0\rangle$ is in the paramagnetic phase \cite{Fradkin:2013aa}. We then apply the decoherence channel $\rho=\prod_{i<j}\mathcal{E}_{ij}[|\psi_0\rangle \langle\psi_0|]$, where
  \begin{equation}
  \mathcal{E}_{ij}[\sigma]=\Big(1-\frac{\mu}{N}\Big)\sigma+\frac{\mu}{N}Z_iZ_j\sigma Z_iZ_j.
  \end{equation}
  Here, the decoherence strength scales as $1/N$, which guarantees a well-defined thermodynamical limit at $N\rightarrow \infty$ \footnote{The reason for choosing all-to-all decoherence is partly because we have numerically verified that local decoherence does not lead to a finite SW-SSB phase region. }. This decoherence is particularly implementable for Rydberg atom arrays, which have all-to-all connectivity \cite{Evered:2023aa,Ma:2023aa,Bluvstein:2024aa,Bekenstein:2020aa,Bluvstein:2021aa,Ebadi:2022aa,Bluvstein:2022aa,Lis:2023aa,Manetsch:2024aa,Tao:2024aa,Cao:2024aa}. 
  

  \begin{figure}[t]
    \centering
    \includegraphics[width=0.9\linewidth]{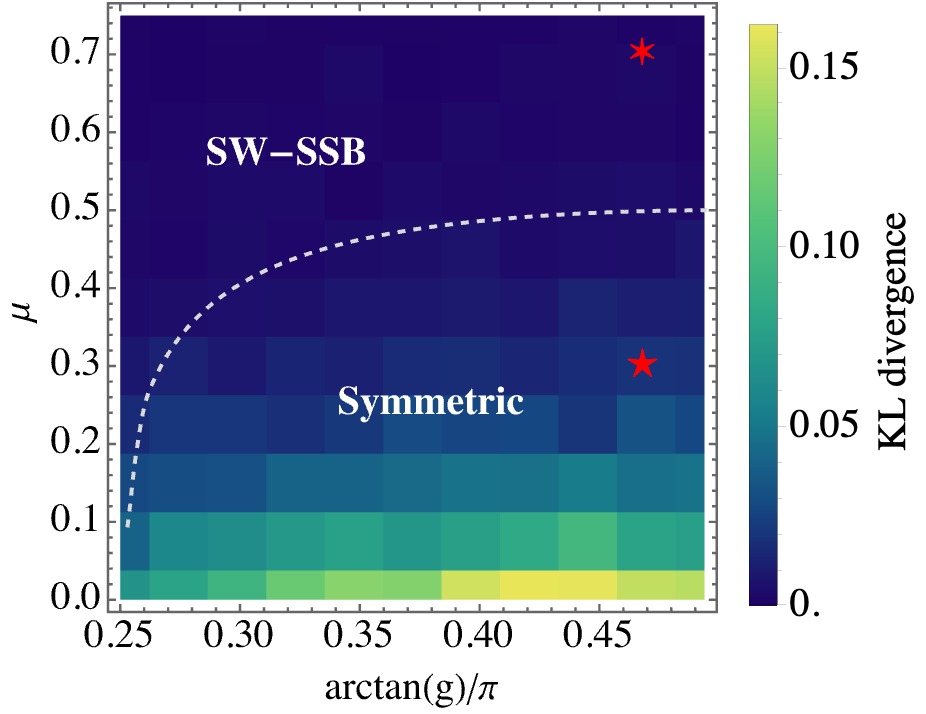}
    \caption{We present the phase diagram of the (all-to-all) decohered quantum Ising model. The phase diagram consists of a symmetric phase at small $\mu$ and a SW-SSB phase at large $\mu$. The white curve represents the theoretical prediction, with the ground-state correlation function obtained from an MPS simulation with $N=200$. The density plot shows the KL divergence between the distribution for $P_I$ and the distribution for $P_{ZZ}$ for a system size of $N=7$ with $N_a=10^4$. The result provides a reasonable estimate of the phase boundary. }
    \label{fig:phase diagram}
  \end{figure}

  We can control the symmetry breaking by tuning parameter $\mu$. At $\mu=0$, $\rho$ is a pure state with a vanishing R\'enyi-2 correlator. As $\mu$ is increased beyond a critical value $\mu_c$, the system undergoes a transition into the phase with SW-SSB. Theoretically, the critical point $\mu_c$ can be related to the correlation function of the un-decohered state $|\psi_0\rangle$. To see that, we employ the Choi-Jamiolkowski isomorphism \cite{Jamiokowski:1972aa,Choi:1975aa} to map the density matrix $\rho$ into a state in the doubled Hilbert space, 
  \begin{equation}
  |\rho\rangle \rangle= \exp\Big(\frac{\mu}{N}\sum_{i<j}(Z_iZ_j\tilde{Z}_i\tilde{Z}_j-1)\Big)|\psi_0\rangle\rangle,
  \end{equation}
  where $|\psi_0\rangle\rangle\equiv |\psi_0\rangle\otimes |\psi_0^*\rangle$. Here we have neglected terms that are subleading in $1/N$. The $\tilde{Z}_i$ with $i=1,2,...,N$ are Pauli operators on the auxiliary Hilbert space. The SW-SSB order can be detected by computing $\overline{C^{(2)}}=\sum_{i<j}C^{(2)}(i,j)/\binom{N}{2}$, which satisfies 
  \begin{equation}\label{eqn:order_parameter}
  \overline{C^{(2)}}=\frac{1}{N-1}\partial_{\mu}\left[e^{\mu N} \langle \langle \rho|\rho\rangle \rangle\right].
  \end{equation}

  The calculation of $\langle \langle \rho|\rho\rangle \rangle$ simplifies significantly in the thermodynamic limit, where the mean-field approximation becomes exact. We first perform the Hubbard-Stratonovich transformation to obtain
  \begin{equation}
  e^{\mu N} \langle \langle \rho|\rho\rangle \rangle=\int \frac{d\phi}{Z_0}~e^{-\mu N \phi^2 } \langle \langle \psi_0 |e^{2\mu\phi\sum_{i}Z_i\tilde{Z}_i}|\psi_0\rangle \rangle,
  \end{equation}
  where $Z_0=\sqrt{\pi/\mu N}$ is a normalization factor. In the supplementary material \cite{SM}, we further provide an exact calculation for arbitrary $N$ in the limit of $g\rightarrow \infty$. Here, we focus on the $N\rightarrow \infty$ and the integration over $\phi$ can be approximated by its saddle-point value $\phi_*$, determined by either $\phi_*=0$ or
  \begin{equation}\label{eqn:saddle}
  \phi_*=\frac{1}{N}\frac{\langle \langle \psi_0 |\sum_{i}Z_i\tilde{Z}_i~e^{2\mu\phi_*\sum_{i}Z_i\tilde{Z}_i}|\psi_0\rangle \rangle}{\langle \langle \psi_0 |e^{2\mu\phi_*\sum_{i}Z_i\tilde{Z}_i}|\psi_0\rangle \rangle}.
  \end{equation}
  Using Eq. \eqref{eqn:order_parameter} and \eqref{eqn:saddle}, one can also show that $\overline{C^{(2)}}=(\phi_*)^2$. Therefore, the manifestation of SW-SSB is indicated by a non-zero saddle-point value, analogous to a traditional Landau paradigm of phase transitions. As a consequence, the phase boundary is given by expanding Eq.~\eqref{eqn:saddle} near $\phi_*=0$, which leads to
  \begin{equation}\label{eqn:muc}
  \mu_c=\Big({2+2\sum_{j\neq 0}\langle \psi_0|Z_jZ_{0}|\psi_0\rangle^2}\Big)^{-1}.
  \end{equation}
  This result applies to arbitrary initial state $|\psi_0\rangle$ with translation symmetry. 

  For the quantum Ising model in Eq.~\eqref{eqn:Ising}, the ground-state correlation functions for any $g$ can be computed using the Jordan-Wigner transformation \cite{Fradkin:2013aa}. In particular, the $ZZ$- correlator can be expressed as a determinant \cite{Pfeuty:1970aa}. Here, 
  we consider several special cases for analytical analysis.
  First, at the critical point $g=1$, the initial state $|\psi_0\rangle$ can be described by Ising CFT, which predicts $\langle \psi_0|Z_jZ_{0}|\psi_0\rangle\propto |j|^{-1/4}$. Thus, the summation in Eq. \eqref{eqn:muc} diverges and the system manifests SW-SSB order for arbitrarily weak decoherence. This reflects the fact that all-to-all decoherence is a relevant perturbation. Second, at $g\rightarrow \infty$, the correlation function vanishes and $\mu_c =1/2$. Third, for large $g$, the correlation length scales as $1/\ln g$ and $\mu_c=(2+\#/\ln g)^{-1}$. Besides, the overall predicted phase boundary for arbitrary $g$ is plotted in Fig. \ref{fig:phase diagram} as a white dashed line. Here, we use correlation functions obtained by a MPS simulation using the \texttt{ITensors.jl} package \cite{Fishman:2022aa} for a system size $N=200$ with open boundary conditions.

  \begin{figure}[t]
    \centering
    \includegraphics[width=0.99\linewidth]{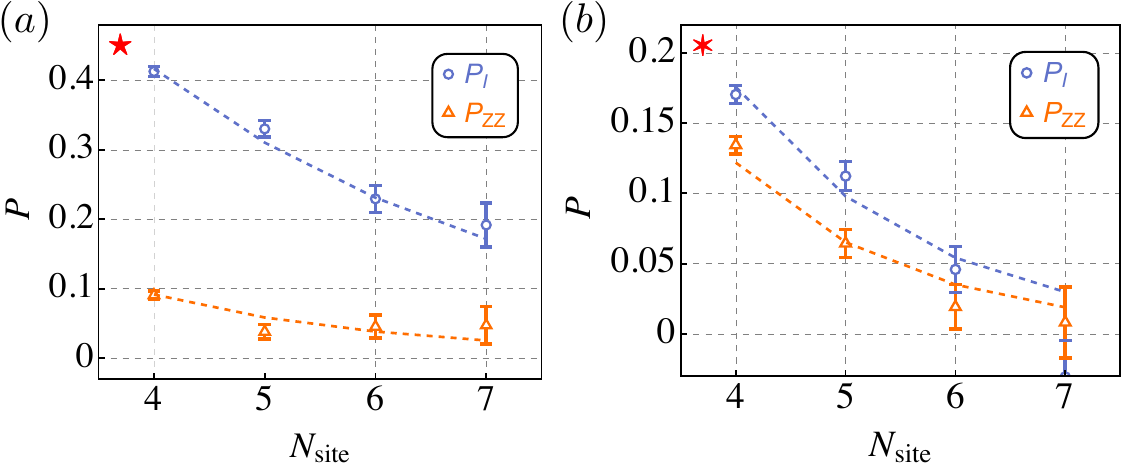}
    \caption{ We present the results of numerical simulations using randomized measurements to predict both the purity $P_I$ and the numerator of the R\'enyi-2 correlator $P_{ZZ}$ for $i=1$ and $j=N$. Here, we choose parameters (a) $(g,\mu)=(10,0.3)$ and (b) $(g,\mu)=(10,0.7)$, corresponding to the star and hexagram in Fig. \ref{fig:phase diagram}, respectively. The result is averaged over $N_a=10^6$ samples, with the error bar representing the standard deviation. The dashed lines denote the exact value predicted by a direct exact diagonalization.}
    \label{fig:randomized_res}
  \end{figure}

  \emph{ \color{blue}Numerical Demonstration.--} 
  We demonstrate our protocol for measuring the R\'enyi-2 correlator using numerical simulations. In our calculation, we assume efficient preparation of the ground state $|\psi_0\rangle$. We implement the all-to-all quantum channel $\prod_{i<j}\mathcal{E}_{ij}$ using quantum trajectory methods without any requirement for non-local operations. In steps 2 and 3 of the experiment for each round, the system is first prepared in $|\psi_0\rangle$. Later for each qubit pair $(i,j)$, we apply $Z_i Z_j$ with a probability of $\mu/N$. The ensemble average over the different choices then yields the target density matrix $\rho$.

  Our numerical simulations are performed using exact diagonalization. We focus on two characteristic parameter sets, $(g,\mu)=(10,0.3)$ in the symmetric phase and $(g,\mu)=(10,0.7)$ in the SW-SSB phase. We numerically repeat $N_a=10^6$ rounds of the experiment with independent measurement directions $\bm{n}$ and quantum trajectories. The collected result from each round contains a pair of bit strings, $\bm{s}$ and $\bm{s}'$. We compute the $F_a = 2^{N} (-2)^{-D(\bm{s}_a,\bm{s}'_a)}$ as the output of the round. The experimental observable is averaged over all the samples following the Eq.~\eqref{eqn:relation}. 
  The statistical error is accounted by dividing the standard deviation of the sample $F_a$ by a factor of $\sqrt{N_a}$.
  The numerical results of the observed $P_{ZZ}$ are shown as data points in FIG. \ref{fig:randomized_res} . For comparison, we also show the observed purity $P_I$. The dashed lines represent the exact theoretical values obtained from Eq.~\ref{def:Renyi-2} with direct exact diagonalization.
  These results clearly demonstrate that our protocol provides an unbiased estimate of $P_{ZZ}$ with a variance comparable to that of the purity $P_{I}$. In particular, for small system sizes $N\lesssim 6$, the protocol already provides a good estimate of the expected value of $P_{ZZ}$. Furthermore, for systems where we can directly apply the quantum channel $\prod_{i<j}\mathcal{E}_{ij}$, it is natural to expect a further decrease in the sample complexity.

\begin{figure}[t]
    \centering
    \includegraphics[width=0.99\linewidth]{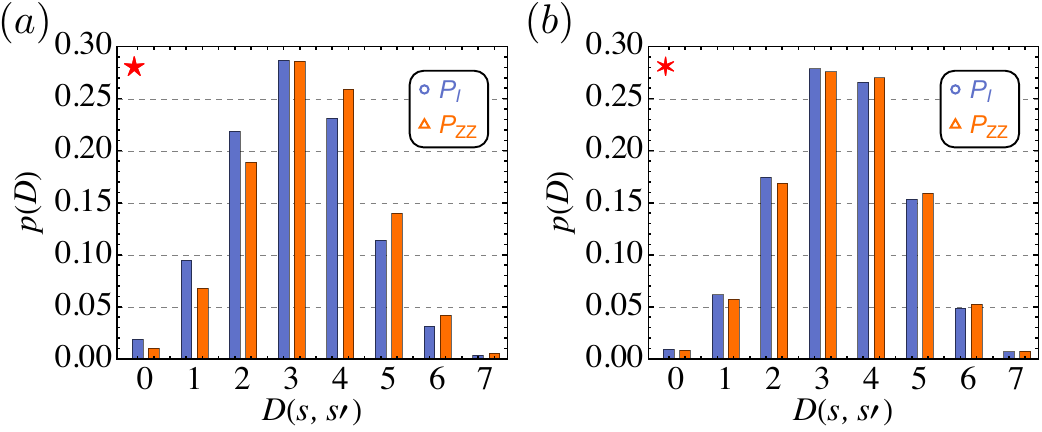}
    \caption{ We present the distribution of the Hamming distance $D(\bm{s},\bm{s}')$ in the numerical simulation with $N_a=10^6$ and $N=7$ for both the purity $P_I$ and the numerator of the R\'enyi-2 correlator $P_{ZZ}$ for $i=1$ and $j=N$. We choose the following parameter sets: (a) $(g,\mu)=(10,0.3)$ and (b) $(g,\mu)=(10,0.7)$, corresponding to the star and hexagram in Fig. \ref{fig:phase diagram}, respectively.}
    \label{fig:randomized_res2}
  \end{figure}

  Our numerical results also shows that at a large system size, such as $N\gtrsim 7$, small sample size is insufficient to obtian meaningful measurement of the R\'enyi-2 correlator.
  In particular, for $(g,\mu)=(10,0.7)$, the standard deviation of both $P_I$ and $P_{ZZ}$ become comparable to their expectations. Thus, it requires a significantly larger $N_a$ to directly estimate $C^{(2)}(i,j)$, which is unfavorable for any realistic experiments. Here, we propose an alternative criterion for detecting the SW-SSB using randomized measurement outcomes in finite-size systems with finite sample size $N_a$, bypassing the calculation of the R\'enyi-2 correlator. The basic observation is that a non-vanishing R\'enyi-2 correlator in the SW-SSB phase requires the expectations $P_{I}$ and $P_{ZZ}$ to be of the same order, such as that in FIG. \ref{fig:randomized_res} (b). This implies that their distributions of Hamming distance $D(\bm{s},\bm{s}')$, denoted as $p_I(D)$ and $p_{ZZ}(D)$, are similar. We demonstrate it by a direct analysis of the numerical data, shown in FIG. \ref{fig:randomized_res2} (b). On the other hand, for systems in the symmetric phase, the expectation value of $P_{ZZ}$ is much smaller than that of $P_I$, shown in FIG. \ref{fig:randomized_res} (a), and their distributions $p_I(D)$ and $p_{ZZ}(D)$ should be distinguishable, shown in FIG. \ref{fig:randomized_res2} (a). Motivated by this observation, we propose to detect the transition using the Kullback-Leibler (KL) divergence 
  \begin{equation}\label{eqn:KL}
  S_{\text{KL}}(p_I|p_{ZZ})=\sum_{D=0}^N p_I(D) \ln\frac{p_I(D)}{p_{ZZ}(D)}.
  \end{equation}
  For the data in FIG. \ref{fig:randomized_res2} (a) and (b), the KL divergence are found to be $1.6\times 10^{-2}$ and $6\times 10^{-4}$, respectively, showing a significantly large gap of two orders of magnitude. Applying this idea to generic $(g,\mu)$ with $N_a=10^4$, we obtain the density plot shown in FIG. \ref{fig:phase diagram}. Amazingly, this simulation, with a small system size $N=7$ and a small number of samples $N_a=10^4$, already provides a reasonable estimate of the phase boundary. We expect that a larger system size in a realistic experiment will lead to further improvement. 

  \emph{ \color{blue}Discussions.--} In this letter, we investigate the application of the randomized measurement toolbox to detect strong-to-weak symmetry breaking of density matrices. Our scheme involves collecting the results of Pauli random measurements for (i) the original quantum state and (ii) the quantum state after the additional application of $Z_iZ_j$. We also propose a relevant experimental model for SW-SSB, which involves all-to-all strongly symmetric decoherence applied to the quantum Ising model. We demonstrate our protocol in this model with a moderate system size and further introduce the KL divergence as an alternative probe of SW-SSB. We anticipate that our results can be readily tested in state-of-the-art quantum devices.
  
  We conclude our work with a few remarks. First, an alternative approach to measuring the second R\'enyi entropy is to prepare two identical systems and perform a measurement of the SWAP operator. In the supplementary material \cite{SM}, we provide an experimental protocol for measuring the R\'enyi-2 correlator following a similar approach. Second, in addition to the R\'enyi-2 correlator, the fidelity correlator \cite{Lessa:2024aa} and Wightman (or R\'enyi-1) correlator \cite{Liu:2024ab,Weinstein:2024aa} have also been introduced as alternative definitions of SW-SSB, which satisfy a series of favorable conditions. It is therefore interesting to ask whether a similar experimental scheme can be proposed to measure Wightman correlators. We defer a careful analysis of this problem to future work.

\vspace{5pt}
\textit{Acknowledgement.} 
We thank Tian-Gang Zhou for helpful discussions. This project is supported by the Shanghai Rising-Star Program under grant number 24QA2700300 (PZ), the NSFC under grant 12374477 (PZ), the Innovation Program for Quantum Science and Technology ZD0220240101 (PZ) and 2023ZD0300900 (LF), and Shanghai Municipal Science and Technology Major Project grant 24DP2600100 (NS and LF).

\bibliography{ref.bib}

\end{document}


\title{Supplementary Material for 
``Scheme to Detect the Strong-to-weak Symmetry Breaking via Randomized Measurements"}

  \author{Ning Sun}
  \affiliation{Department of Physics, Fudan University, Shanghai, 200438, China}

  \author{Pengfei Zhang}
  \thanks{PengfeiZhang.physics@gmail.com}
  \affiliation{Department of Physics, Fudan University, Shanghai, 200438, China}
  \affiliation{State Key Laboratory of Surface Physics, Fudan University, Shanghai, 200438, China}
  \affiliation{Shanghai Qi Zhi Institute, AI Tower, Xuhui District, Shanghai 200232, China}
  \affiliation{Hefei National Laboratory, Hefei 230088, China}

  \author{Lei Feng}
  \thanks{leifeng@fudan.edu.cn}
  \affiliation{Department of Physics, Fudan University, Shanghai, 200438, China}
  \affiliation{State Key Laboratory of Surface Physics, Fudan University, Shanghai, 200438, China}
  \affiliation{Institute for Nanoelectronic devices and Quantum computing, Fudan University, Shanghai, 200438, China}
  \affiliation{Shanghai Key Laboratory of Metasurfaces for Light Manipulation, Shanghai, 200433, China}
  \affiliation{Hefei National Laboratory, Hefei 230088, China}
  
  \date{\today}
\maketitle

In this supplementary material, we present (1) an exact formula for $\overline{C^{(2)}}$ in the limit $g\rightarrow \infty$; (2) an alternative experimental scheme utilizing two copies of the system; (3) a brief review of the Kullback-Leibler (KL) divergence.

\section{Exact results in the limit \texorpdfstring{$g \rightarrow \infty$}{TEXT} for arbitrary \texorpdfstring{$N$}{TEXT} }

In the main text, we focus on the thermodynamic limit with \(N \to \infty\). Here, we consider a different limit \(g \to \infty\), where an analytical solution of \(\overline{C^{(2)}}\) is possible for arbitrary \(N\). We begin by keeping Eq. (8) in the main text to arbitrary order in \(1/N\): 
\begin{equation}
|\rho\rangle \rangle = \left[\frac{1-\mu/N}{\cosh u}\right]^{\frac{N(N-1)}{2}} e^{-\frac{uN}{2}} e^{\frac{u}{2} \sum_{ij} Z_i Z_j \tilde{Z}_i \tilde{Z}_j} |\psi_0\rangle \rangle 
\equiv \left[\frac{1-\mu/N}{\cosh u}\right]^{\frac{N(N-1)}{2}} |\rho_u\rangle \rangle.
\end{equation}

Here, we have introduced the effective coupling \(u\) satisfying \(\tanh u = \frac{\mu}{N-\mu}\) (recall that we require \(\mu \in [0, N]\) for finite \(N\)). To proceed, it is more convenient to express the averaged R\'enyi-2 correlator as 
\begin{equation}
\overline{C^{(2)}} = \frac{1}{N(N-1)} \partial_u \left[\ln \langle \langle \rho_u|\rho_u\rangle \rangle \right].
\end{equation}

Next, we perform the Hubbard-Stratonovich transformation, which leads to 
\begin{equation}
\begin{aligned}
\langle \langle \rho_u|\rho_u\rangle \rangle =& \sqrt{\frac{u}{\pi}} e^{-u N} \int_{-\infty}^\infty d\Phi~e^{-u \Phi^2} \langle \langle \psi_0|e^{2u \Phi \sum_i Z_i \tilde{Z}_i}|\psi_0\rangle \rangle \\
=& \sqrt{\frac{u}{\pi}} e^{-u N} \int_{-\infty}^\infty d\Phi~e^{-u \Phi^2} \left[\cosh(2u \Phi)\right]^N 
= \frac{e^{-u N}}{2^N} \sum_{m=0}^N \binom{N}{m} e^{u (N-2m)^2}.
\end{aligned}
\end{equation}

This leads to the final result:
\begin{equation}
\overline{C^{(2)}} = \frac{1}{N(N-1)} \left[-N + \frac{\sum_{m=0}^N \binom{N}{m} (N-2m)^2 e^{u (N-2m)^2}}{\sum_{m=0}^N \binom{N}{m} e^{u (N-2m)^2}} \right].
\end{equation}

\section{Alternative Scheme}

We propose an alternative scheme to detect strong-to-weak spontaneous symmetry breaking (SW-SSB) through interference measurements on duplicated systems. The purity, $\tr[\rho^2]$, and consequently the \renyitwo entropy, has already been successfully measured in experiments utilizing interference on duplicated systems \cite{Islam:2015aa,Kaufman:2016aa}. Inspired by these achievements, we suggest that the \renyitwo correlator can also be experimentally measured using a similar setup.

Recall that the \renyitwo correlator is defined as  
\begin{equation}\label{def:Renyi-2}  
  C^{(2)}(j,k) = \frac{\tr[O_j O_k^\dag \rho O_k O_j^\dag \rho]}{\tr[\rho^2]}.  
\end{equation}  
The denominator, \(\tr[\rho^2]\), represents the purity and can be straightforwardly estimated by measuring the swap operation on duplicated systems \cite{Islam:2015aa,Kaufman:2016aa}. The numerator, \(\tr[O_j O_k^\dag \rho O_k O_j^\dag \rho]\), can similarly be estimated by measuring the swap operation on duplicated systems, with the additional application of the operator \(O_j O_k^\dag\) on one of the duplicates. Specifically, this requires measuring the swap operation between \(\rho\) and \(\tilde{\rho}\), where \(\tilde{\rho}\) is a copy of \(\rho\) with the operator \(O_j O_k^\dag\) applied.  

Formally, these quantities can be expressed as  
\begin{equation}  
    \tr[\rho^2] = \langle V \rho \otimes \rho \rangle,  
\end{equation}  
and  
\begin{equation}  
    \tr[O_j O_k^\dag \rho O_k O_j^\dag \rho] = \langle V \rho \otimes \tilde{\rho} \rangle,  
\end{equation}  
with  
\begin{equation}  
    \tilde{\rho} = O_j O_k^\dag \rho O_k O_j^\dag.  
\end{equation}  
Here, \(V\) denotes the swap operation between the two duplicated systems. The derivation is presented as follows.

\begin{proof}
Since $V$ is the swap operator defined in the tensor-product space, we have
\begin{equation}
    V|m\rangle|n\rangle = |n\rangle|m\rangle. 
\end{equation}
For general operator $A$ and $B$, we have

\begin{equation}
\begin{aligned}
    A\otimes B &= \sum_{m,m',n,n'}(A_{m'n'}|m'\rangle\langle n'|)\otimes(B_{mn}|m\rangle\langle n|) \\
    &= \sum_{m,m',n,n'}A_{m'n'}B_{mn}(|m'\rangle\otimes|m\rangle)(\langle n'|\otimes\langle n|) \\
    &= \sum_{m,m',n,n'}A_{m'n'}B_{mn}|m'\rangle|m\rangle\langle n'|\langle n|. 
\end{aligned}
\end{equation}
Then,
\begin{equation}
\begin{aligned}
    VA\otimes B &= V  \sum_{m,m',n,n'}A_{m'n'}B_{mn}|m'\rangle|m\rangle\langle n'|\langle n| \\
    &= \sum_{m,m',n,n'}A_{m'n'}B_{mn}V|m'\rangle|m\rangle\langle n'|\langle n| \\
    &=  \sum_{m,m',n,n'}A_{m'n'}B_{mn}|m\rangle|m'\rangle\langle n'|\langle n|. 
\end{aligned}
\end{equation}
Therefore, 
\begin{equation}
\begin{aligned}
    \tr(VA\otimes B) &= 
    \tr(\sum_{m,m',n,n'}A_{m'n'}B_{mn}|m\rangle|m'\rangle\langle n'|\langle n|) \\
    &= \sum_{kl}\langle k|\langle l| \left(\sum_{m,m',n,n'}A_{m'n'}B_{mn}|m\rangle|m'\rangle\langle n'|\langle n|\right) |k\rangle|l\rangle \\
    &= \sum_{kl}\sum_{m,m',n,n'}\delta_{km}\delta_{lm'}\delta_{n'k}\delta_{nl}A_{m'n'}B_{mn} \\
    &= \sum_{kl}A_{lk}B_{kl} \\
    &= \tr(AB). 
\end{aligned}
\end{equation}
As two concrete examples, we obtain
\begin{equation}
    \tr[\rho^2] = \tr[V\rho\otimes\rho],
\end{equation}
and
\begin{equation}
    \tr[O_jO_k^\dag\rho O_kO_j^\dag\rho] = \tr[V\rho\otimes\tilde\rho], 
\end{equation}
with
\begin{equation}
    \tilde\rho = O_jO_k^\dag\rho O_kO_j^\dag. 
\end{equation}
\end{proof}

The remaining task is to measure the expectation value of the swap operator on the two duplicated systems. This can be accomplished using well-established techniques involving quantum many-body interference and site-resolved measurements on optical lattices \cite{Islam:2015aa,Kaufman:2016aa}.

\section{On the Kullback-Leibler divergence}

The Kullback-Leibler (KL) divergence, also called KL distance, is an effective method for characterizing the relative entropy of one distribution $P(x)$ with respect to another $Q(x)$, defined as
\begin{equation}
    S_{\kl}(P || Q) = \int \dx P(x)\log\left(\dfrac{P(x)}{Q(x)}\right).
\end{equation}
It is not symmetric between $P$ and $Q$, since $S_{\kl}(P || Q) \neq S_{\kl}(Q || P)$ in general. 
The minimal value of the KL divergence is 0, reached at $P = Q$. 
We demonstrate this as follows. 

\begin{proof}
\begin{equation}
    S_{\kl}(P||Q) = \int \dx P(x)\log\left(\dfrac{P(x)}{Q(x)}\right) = \int\dx P(x) \log P(x) - \int\dx P(x) \log Q(x),
\end{equation}
Taking \(P(x)\) as given, the first term in the KL divergence becomes a constant.  
Only the second term varies as \(Q(x)\) changes.  
Thus, to minimize \(S_{\kl}(P || Q)\), we focus on minimizing  
\begin{equation}\label{eq:dq}  
    -\int \dx P(x)\log Q(x),  
\end{equation}  
subject to the constraint  
\begin{equation}  
    \int Q(x) \dx = 1,  
\end{equation}  
since a probability distribution must integrate to unity.  

The constraint can be incorporated by introducing a Lagrange multiplier term.  
Therefore, we construct the following function \(f(Q, \lambda)\) to minimize:  
\begin{equation}  
    f(Q, \lambda) = -\int P(x)\log Q(x) \dx + \lambda\left(\int Q(x)\dx - 1\right).  
\end{equation}  

The saddle point of \(f(Q, \lambda)\) can be obtained by performing a variation with respect to \(Q(x)\) and taking the partial derivative with respect to \(\lambda\), then setting both to zero.  
\begin{equation}\label{eq:saddle}  
\begin{aligned}  
    \delta f &= \int \left(-P(x) \dfrac{1}{Q(x)} + \lambda\right)\delta Q(x) \dx = 0, \\  
    \dfrac{\partial f}{\partial \lambda} &= \int Q(x)\dx - 1 = 0.  
\end{aligned}  
\end{equation}  

The variation \(\delta Q(x)\) is independent for different values of \(x\).  
To ensure the first term sums to zero, it must vanish for all \(x\), leading to the condition  
\begin{equation}  
    -\dfrac{P(x)}{Q(x)} + \lambda = 0  
\end{equation}  
for all \(x\).  
This implies that \(Q(x)\) must be proportional to \(P(x)\):  
\begin{equation}  
    Q(x) = \dfrac{1}{\lambda} P(x).  
\end{equation}  

Substituting this into the second line of \eqref{eq:saddle}, we find  
\begin{equation}  
    \int \dfrac{1}{\lambda} P(x)\dx - 1 = 0.  
\end{equation}  

Since \(P(x)\) is a probability distribution, it satisfies \(\int P(x)\dx = 1\).  
Thus, solving for \(\lambda\), we obtain  
\begin{equation}  
    \lambda = 1.  
\end{equation}  

Finally, substituting \(\lambda = 1\) back, we find that the saddle point of the KL divergence \(S_{\kl}(P || Q)\) is achieved when  
\begin{equation}  
    Q(x) = P(x).  
\end{equation}  

Expanding \eqref{eq:dq} to the second order, we have  
\begin{equation}  
    -\int P(x) \log(Q + \delta Q) \dx  
    = \int \dx \left[-P(x) \left(\log Q + \dfrac{\delta Q}{Q}  
    - \dfrac{\delta Q^2}{2Q^2} + \mathcal{O}(\delta Q^3)\right)\right].  
\end{equation}  

The coefficient of the second-order variation \(\delta Q^2\) is  
\begin{equation}  
    \dfrac{P}{2Q^2},  
\end{equation}  
which is always non-negative.  

Thus, it is confirmed that \(Q = P\) corresponds to a minimum of \(S_{\kl}(P || Q)\).

\end{proof}

\bibliography{ref.bib}